\documentclass[pdflatex,sn-vancouver-num]{sn-jnl}
\usepackage{graphicx}%
\usepackage{multirow}%
\usepackage{amsmath,amssymb,amsfonts}%
\usepackage{amsthm}%
\usepackage{mathrsfs}%
\usepackage[title]{appendix}%
\usepackage{xcolor}%
\usepackage{textcomp}%
\usepackage{manyfoot}%
\usepackage{booktabs}%
\usepackage{algorithm}%
\usepackage{algorithmicx}%
\usepackage{algpseudocode}%
\usepackage{listings}%
\usepackage{braket}

\theoremstyle{thmstyleone}%
\theoremstyle{thmstyletwo}%

\theoremstyle{thmstylethree}%

\begin{document}

\title{An Entanglement-Assisted Stabilizer Framework for Distributed Sensing of Local Phases}


\author[1]{\fnm{Huidan} \sur{Zheng}}%\email{hyedan@korea.ac.kr}

\author*[2]{\fnm{Ilkwon} \sur{Sohn}}\email{d2estiny@kisti.re.kr}

\author*[1]{\fnm{Jun} \sur{Heo}}\email{junheo@korea.ac.kr}

\affil*[1]{\orgdiv{School of Electrical Engineering}, \orgname{Korea University}, \orgaddress{\city{Seoul}, \postcode{02841}, \country{Republic of Korea}}}

\affil[2]{\orgdiv{Quantum Network Research Center}, \orgname{Korea Institute of Science and Technology Information}, \orgaddress{\city{Daejeon}, \postcode{34141}, \country{Republic of Korea}}}


\abstract{Distributed quantum sensing requires spatially separated probes to acquire local parameters while maintaining compatibility with network-level quantum information processing. We develop an entanglement-assisted stabilizer framework based on the extended structure of entanglement-assisted quantum error-correcting (EAQEC) codes, in which the remote halves of pre-shared ebits are used directly as local phase probes while the joint state simultaneously carries an encoded logical subsystem. Each remote probe acquires a local $Z$-axis phase and subsequently returns through an $X$-type noise channel. Within the extended EAQEC stabilizer structure, the stabilizer containing $X_{B_j}$ provides phase-dependent measurement statistics, whereas its partner containing $Z_{B_j}$ records the corresponding return-error syndrome. A graph-code formulation is introduced to make this structure explicit, together with an illustrative $[[5,1,3;2]]$ construction. We further show that, conditioned on the joint sensing-and-syndrome measurement record, the post-sensing state differs from the original encoded state only by a known Pauli transformation, so that the logical information remains available for subsequent encoded operations. For the local-phase model considered here, the stabilizer readout attains the available quantum Fisher information, while the finite-shot estimator approaches the corresponding $1/\sqrt{M}$ scaling as the number of repetitions increases.
The framework therefore provides a common entanglement-assisted stabilizer structure for distributed local-phase sensing, restricted return-error identification, and post-sensing logical-state retention, without relying on an intrinsic metrological enhancement from EAQEC itself.}




\maketitle

\section{Introduction}\label{sec1}

Quantum metrology exploits quantum coherence and entanglement to estimate physical parameters with precision beyond classical measurement strategies \cite{Giovannetti2011}. As quantum sensing extends from individual probes to spatially distributed sensor networks, the parameters of interest may be encoded locally at different sensing nodes and inferred through measurements performed across the network. Such settings arise naturally in multiparameter and distributed sensing, where spatially separated probes interact with different local parameters and their correlations can be used to estimate either individual parameters or functions of them \cite{Proctor2018,Eldredge2018,Kim2024,Bate2025}.

Distributed sensing can also separate local signal acquisition from quantum information processing. A remote probe need only interact with its local environment and subsequently participate in the readout, while state preparation, encoding, and collective measurement may be concentrated elsewhere in the network~\cite{Zhang2026}. This separation motivates sensing architectures in which remote qubits act as local probes without requiring an independent error-correcting processor at every sensing location. In this setting, however, the interaction experienced by a probe cannot be classified solely as noise: the parameter-dependent evolution is the signal to be retained, whereas other perturbations may still need to be detected or corrected.

This distinction is central to quantum-error-corrected metrology. Quantum error correction (QEC) has been used to extend sensing coherence and recover metrological performance in noisy environments, provided that the error-correcting procedure suppresses the relevant noise without removing the signal Hamiltonian \cite{Arrad2014,Kessler2014,Dur2014}. Early proposals demonstrated this principle in several sensing settings, and subsequent work established general conditions under which QEC can recover Heisenberg-limited scaling under Markovian noise \cite{Zhou2018}, with recent work extending QEC-metrological analyses to non-Markovian noise models \cite{Mann2025}. Ancilla-free constructions have further shown that noiseless ancillary systems are not always necessary \cite{Layden2019,Zhou2024}. More recently, the tension between signal sensitivity and error protection has been formulated directly as a protection-sensitivity incompatibility, with asymmetric codes proposed to relax protection along the signal direction while retaining protection against complementary perturbations \cite{Chen2025}.

Syndrome information can likewise play a role beyond conventional recovery. Stabilizer syndromes obtained during QEC can contain information about underlying Pauli noise channels while leaving the encoded quantum state intact \cite{Wagner2022}. Recent encoded quantum signal processing approaches have gone further by using syndrome measurements themselves as signal-processing primitives for logical sensors, with the objective of retaining signal sensitivity while suppressing noise \cite{OrtizMarrero2026}. These developments establish that syndrome measurements need not be viewed solely as indicators of errors; their statistics can also encode information about physical processes acting on an encoded system.

Entanglement-assisted quantum error correction (EAQEC) provides a distinct setting in which this idea can be explored in a distributed architecture. An $[[n,k,d;c]]$ EAQEC code uses $c$ pre-shared ebits between the encoder and receiver, with the receiver-side halves entering the joint stabilizer structure although they are not acted upon by the encoder \cite{Brun2006,Hsieh2007}. Conventional EAQEC assumes these receiver-side qubits to be noiseless, while extensions with imperfect ebits treat their errors as additional faults to be corrected \cite{LaiBrun2012}. Separately, distributed sensing protocols have investigated the use of shared entanglement and error-correcting codes to protect metrological performance across sensor networks \cite{Zhuang2020,ZhouBradyZhuang2022}. The setting considered here differs in that the remote halves of the EAQEC ebits themselves are assigned the role of local sensing probes while the same joint state continues to carry an independent encoded logical subsystem.

In this work, each remote ebit half $B_j$ interacts with a local phase $\theta_j$ through a $Z$-axis rotation and subsequently returns through an $X$-type noise channel. The extended EAQEC stabilizer pair associated with $B_j$ then acquires two distinct operational roles: the stabilizer containing $X_{B_j}$ provides phase-dependent measurement statistics, whereas the partner containing $Z_{B_j}$ records the corresponding return-error syndrome. A graph-code representation is used to make this structure explicit and to construct a $[[5,1,3;2]]$ example. The joint measurement record of $\{g_j^X, g_j^Z\}^c_{j=1}$ also determines the conditional post-sensing encoded state up to a known Pauli transformation, allowing the logical information to remain available without direct logical measurement. For the local-phase model considered here, the resulting stabilizer readout attains the available quantum Fisher information, matching the corresponding single-qubit phase-sensing benchmark. The contribution is an entanglement-assisted stabilizer framework that combines distributed local-phase estimation, restricted return-error identification, and post-sensing logical-state retention within the same encoded structure.

\section{Entanglement-assisted stabilizer structure}\label{sec2}

\subsection{Extended EAQEC stabilizer structure}
\label{sec2-1}

We consider an entanglement-assisted quantum error-correcting code with parameters $[[n,k,d;c]]$, where $k$ logical qubits are encoded into an $n$-qubit register held by Alice with the assistance of $c$ pre-shared ebits between Alice and Bob. Alice applies the encoding operation only to her local qubits, while Bob's halves of the ebits, denoted by $B_1,\ldots,B_c$, are not acted upon by the encoder \cite{Brun2006,Hsieh2007}. The encoded state is therefore supported on the joint Alice--Bob system and contains $k$ logical degrees of freedom.

Because Alice's encoding acts locally, the reduced state of Bob's register remains maximally mixed for any logical input,
\begin{equation}
\rho_B
=
\operatorname{Tr}_A(\rho_{\mathrm{EA}})
=
\frac{I_B}{2^c}.
\label{eq:bob-marginal}
\end{equation}
Thus, the remote register alone carries no information about the encoded logical state. This property allows the Bob-side qubits to participate in local sensing interactions without directly exposing the logical information stored in the joint encoded state.

The entanglement-assisted stabilizer structure contains, for each shared ebit, a pair of joint operators of the form
\begin{equation}
g_j^X=P_j\otimes X_{B_j},
\qquad
g_j^Z=Q_j\otimes Z_{B_j},
\label{eq:extended-pair}
\end{equation} where $P_j$ and $Q_j$ are Pauli operators acting on Alice's encoded register.
Their Alice-side restrictions satisfy $\{P_j,Q_j\}=0$, while $X_{B_j}$ and $Z_{B_j}$ also anticommute. The two anticommutations therefore cancel in the joint system, giving
\begin{equation}
[g_j^X,g_j^Z]=0.
\end{equation}
Together with the Alice-only stabilizer generators, these operators define a commuting stabilizer group on the extended Alice--Bob system. The Alice-side restrictions of the remote pair need not commute by themselves; their compatibility is obtained through the corresponding Bob-side Pauli operators, which is the characteristic entanglement-assisted structure used throughout this work.

In the conventional EAQEC setting, the distance $d$ characterizes correctable errors acting on Alice's $n$-qubit encoded register, while Bob's halves of the pre-shared ebits are assumed noiseless \cite{Brun2006,Hsieh2007}. The present sensing framework instead assigns these Bob-side qubits an active role as remote probes. Errors acting on them therefore lie outside the conventional distance guarantee and must be treated explicitly in the sensing and return-channel model.

\subsection{Graph-code realization}\label{sec2-2}

A graph-code representation provides a convenient realization of these remote stabilizer pairs. We partition the graph vertices as $V=I\sqcup O_A\sqcup O_B$, where $I$ contains the $k$ logical input vertices, $O_A=\{A_1,\ldots,A_n\}$ represents Alice's encoded register, and $O_B=\{B_1,\ldots,B_c\}$ represents the remote halves of the pre-shared ebits. The input vertices specify the encoding and are not part of the physical output register, whereas $O_A\sqcup O_B$ describes the joint Alice--Bob system after encoding \cite{SchlingemannWerner2001,Schlingemann2002}.

For the sensing architecture considered here, we restrict attention to graphs in which each remote vertex $B_j$ is a leaf connected to a distinct Alice-side output vertex $a_j$,
\begin{equation}
N_G(B_j)=\{a_j\},
\qquad
a_j\neq a_\ell\quad (j\neq \ell).
\label{eq:remote-leaf}
\end{equation}
We refer to this as the remote-leaf-matched condition. Since $B_j$ has no input neighbor, its output-graph vertex operator belongs to the graph-code stabilizer and takes the form
\begin{equation}
g_j^X=K_{B_j}=Z_{a_j}X_{B_j}.
\label{eq:remote-gx}
\end{equation}
Thus, the remote-leaf topology directly provides the $X_{B_j}$-containing member of the entanglement-assisted stabilizer pair. The complete graph-code characterization underlying this result is given in Appendix~A.

The remote-leaf condition alone, however, does not guarantee a complete entanglement-assisted pair. For each $B_j$, the stabilizer group must also contain an independent operator $g_j^Z=Q_j\otimes Z_{B_j}$, where $Q_j$ acts on Alice's output register and satisfies $\{Z_{a_j},Q_j\}=0$. The anticommutation between $Z_{a_j}$ and $Q_j$ on Alice's side is compensated by that between $X_{B_j}$ and $Z_{B_j}$, so $g_j^X$ and $g_j^Z$ commute on the joint system. These two operators constitute the remote stabilizer pair used in the sensing protocol.

\begin{figure}[t]
    \centering
    \includegraphics[width=0.33\linewidth]{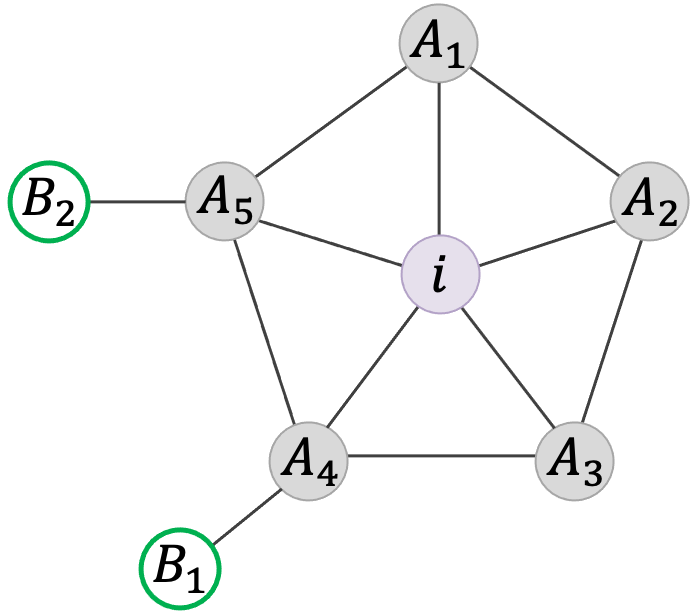}
    \caption{Illustrative $[[5,1,3;2]]$ entanglement-assisted graph-code realization. The five Alice-side output vertices form a cycle, while $B_1$ and $B_2$ are remote leaves attached to $A_4$ and $A_5$, respectively. The input vertex $i$ specifies the logical input and is not part of the physical output register.}
    \label{fig:example}
\end{figure}

An explicit realization is provided by the $[[5,1,3;2]]$ graph code shown in Fig.~\ref{fig:example}. The five Alice-side output vertices form a cycle, while $B_1$ and $B_2$ are remote leaves attached to $A_4$ and $A_5$, respectively. The corresponding remote-leaf operators are
\begin{equation}
g_1^X=Z_{A_4}X_{B_1},
\qquad
g_2^X=Z_{A_5}X_{B_2},
\label{eq:example-gx}
\end{equation}
and compatible $Z$-type partners can be chosen as
\begin{align}
g_1^Z &=
X_{A_1}Z_{A_2}Z_{A_3}X_{A_4}Z_{B_1},
\nonumber\\
g_2^Z &=
X_{A_2}Z_{A_3}Z_{A_4}X_{A_5}Z_{B_2}.
\label{eq:example-gz}
\end{align}
Together with two Alice-only stabilizer generators, these remote pairs realize the $[[5,1,3;2]]$ entanglement-assisted graph code. The complete generating set, logical Pauli representatives, and verification of the conventional EAQEC distance $d=3$ are given in Appendix~B. The $[[5,1,3;2]]$ code serves only as an explicit realization; the sensing framework applies to the general $[[n,k,d;c]]$ structure.

\section{Distributed sensing protocol}\label{sec3}

\subsection{Problem setting and protocol overview}
\label{sec3-1}

Let $B_1,\ldots,B_c$ denote the remote halves of the pre-shared ebits. Each $B_j$ serves as a local probe for a spatially dependent phase parameter $\theta_j$, while Alice retains the $n$-qubit encoded register throughout the sensing stage. We collect the local parameters as $\boldsymbol{\theta}=(\theta_1,\ldots,\theta_c)$. For each probe $B_j$, the parameter $\theta_j$ is restricted to an identifiable operating interval $\mathcal{I}_j$.

The protocol consists of sequential signal acquisition, probe return, and joint stabilizer readout. After preparation of the entanglement-assisted encoded state, each remote probe $B_j$ undergoes a local $Z$-axis phase rotation determined by $\theta_j$, while Alice's encoded register is isolated from the signal. The probes are then returned to Alice through a noisy quantum channel. In the model considered here, signal acquisition and return noise are treated as separate stages, with independent $X$-type errors acting only during the return stage.

\begin{figure}[htb]
\centering
\includegraphics[width=1\textwidth]{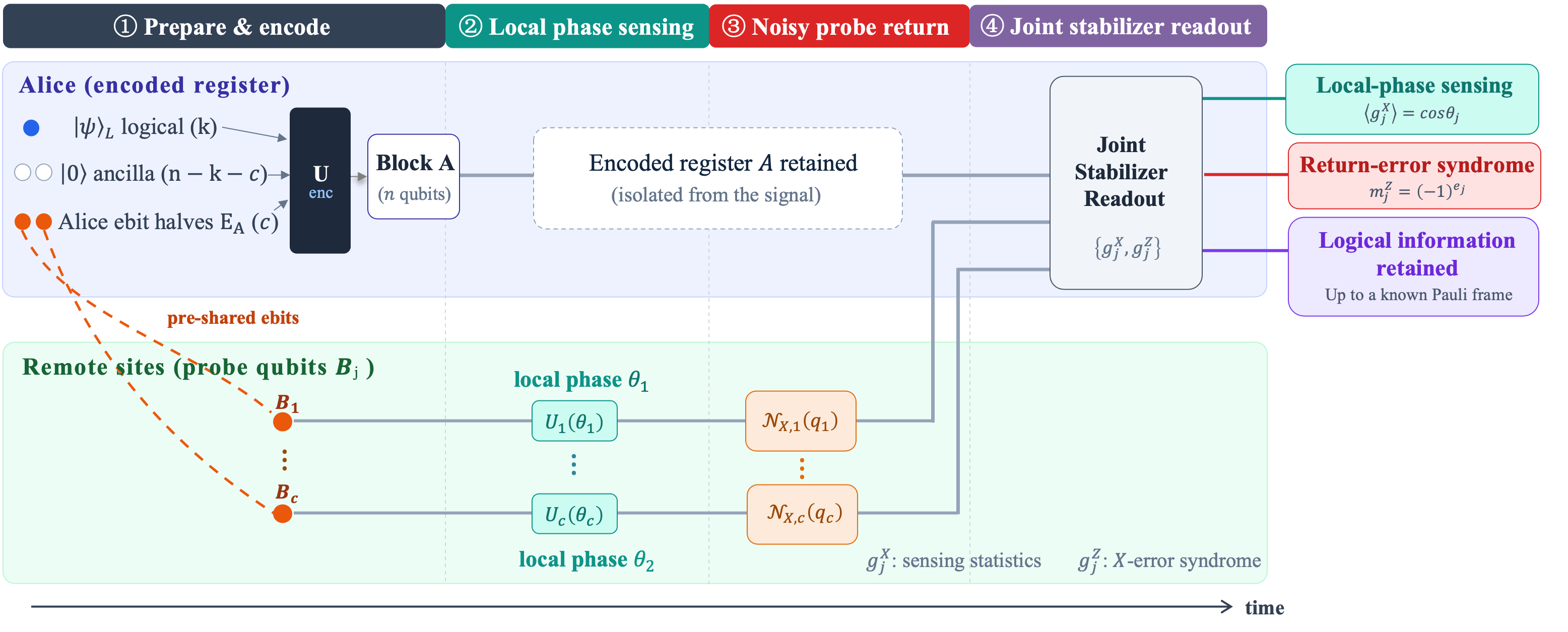}
\caption{Distributed sensing protocol based on the entanglement-assisted stabilizer structure. The remote halves of the pre-shared ebits acquire local phases and return through independent $X$-error channels. Joint stabilizer readout provides phase-dependent sensing statistics and return-error syndromes, while the encoded logical subsystem remains available as a conditional post-sensing output.}
\label{fig:protocol}
\end{figure}

After the probes return, Alice performs a joint readout of the remote stabilizer pairs. The outcomes of $g_j^X$ provide the phase-dependent statistics used to estimate $\theta_j$, whereas the corresponding $g_j^Z$ outcomes identify $X$-type errors acquired during probe return. The logical subsystem is not directly measured in this procedure. The resulting post-sensing encoded state is therefore retained as a conditional quantum output. An overview of the complete protocol is shown in Fig.~\ref{fig:protocol}.

\subsection{Local-phase signal and sensing statistics}\label{sec3-2}

Let $\lvert\Psi_{\mathrm{EA}}(\psi)\rangle$ denote an arbitrary encoded state in the entanglement-assisted code space, satisfying $g_j^X\lvert\Psi_{\mathrm{EA}}(\psi)\rangle
=
g_j^Z\lvert\Psi_{\mathrm{EA}}(\psi)\rangle
=
\lvert\Psi_{\mathrm{EA}}(\psi)\rangle$ for each remote stabilizer pair.

Each remote probe $B_j$ acquires a local phase $\theta_j$ through $U_j(\theta_j)=\exp(-i\theta_j Z_{B_j}/2)$. The joint sensing operation and the resulting state are
\begin{equation}
U_{\boldsymbol{\theta}}^{(B)}
=
\bigotimes_{j=1}^{c}U_j(\theta_j),
\qquad
\lvert\Psi_{\boldsymbol{\theta}}(\psi)\rangle
=
U_{\boldsymbol{\theta}}^{(B)}
\lvert\Psi_{\mathrm{EA}}(\psi)\rangle .
\label{eq:sensing-unitary}
\end{equation}

The remote register remains maximally mixed under these local rotations. The phase information is therefore carried by the modified Alice--Bob correlations rather than by the reduced state of an individual probe.

For a fixed probe $B_j$, define $\lvert\Phi_j\rangle
=
\prod_{\ell\neq j}
U_\ell(\theta_\ell)
\lvert\Psi_{\mathrm{EA}}(\psi)\rangle$. Since the rotations on the other probes commute with $g_j^X$, $g_j^X\lvert\Phi_j\rangle=\lvert\Phi_j\rangle$. The sensed state can then be written as
\begin{equation}
\lvert\Psi_{\boldsymbol{\theta}}(\psi)\rangle
=
\cos\frac{\theta_j}{2}\lvert\Phi_j\rangle
-
i\sin\frac{\theta_j}{2}
Z_{B_j}\lvert\Phi_j\rangle .
\label{eq:sensed-decomposition}
\end{equation}

Because $Z_{B_j}$ anticommutes with the $X_{B_j}$ factor of $g_j^X$, the two terms belong to the $+1$ and $-1$ eigenspaces of $g_j^X$, respectively. The corresponding measurement statistics are
\begin{equation}
p_j^{(\pm)}
=
\frac{1\pm\cos\theta_j}{2},
\qquad
\langle g_j^X\rangle_{\boldsymbol{\theta}}
=
\cos\theta_j .
\label{eq:sensing-statistics}
\end{equation}
The marginal statistics of $g_j^X$ depend only on the corresponding local parameter $\theta_j$, allowing the local phases to be estimated independently.

By contrast, $g_j^Z=Q_j\otimes Z_{B_j}$ commutes with $U_{\boldsymbol{\theta}}^{(B)}$, so ideal phase accumulation leaves its stabilizer eigenvalue unchanged. The two members of each remote stabilizer pair therefore provide phase-sensitive and signal-insensitive readout, respectively.

A single $g_j^X$ measurement depends on $\theta_j$ only through $\cos\theta_j$, so an unrestricted phase cannot be identified uniquely. Choosing the identifiable interval $\mathcal{I}_j=[0,\pi]$ gives
\begin{equation}
\theta_j
=
\arccos\!\left(2p_j^{(+)}-1\right).
\label{eq:phase-inverse}
\end{equation}

\subsection{Probe return and $X$-error syndrome}\label{sec3-3}

After phase acquisition, the remote probes return through independent bit-flip channels,
\begin{equation}
\mathcal{N}_X^{(B)}
=
\bigotimes_{j=1}^{c}\mathcal{N}_{X,j},
\qquad
\mathcal{N}_{X,j}(\rho)
=
(1-q_j)\rho
+
q_j X_{B_j}\rho X_{B_j},
\label{eq:return-channel}
\end{equation}
where $q_j$ is the probability of an $X$ error on probe $B_j$.

For an error pattern $\boldsymbol{e}=(e_1,\ldots,e_c)\in\mathbb{F}_2^c$, write $X_B^{\boldsymbol{e}}
=
\prod_{j=1}^{c}X_{B_j}^{e_j}$. The two members of the remote stabilizer pair respond differently to this error:
\begin{equation}
[X_B^{\boldsymbol{e}},g_j^X]=0,
\qquad
g_j^Z X_B^{\boldsymbol{e}}
=
(-1)^{e_j}
X_B^{\boldsymbol{e}}g_j^Z .
\label{eq:return-commutation}
\end{equation}

Since phase acquisition leaves the $g_j^Z$ eigenvalue unchanged, the outcome of the $g_j^Z$ measurement after a realized return-error pattern is
\begin{equation}
m_j^Z=(-1)^{e_j},
\qquad
\Pr(m_j^Z=-1)=q_j .
\label{eq:return-syndrome}
\end{equation}
Thus, under ideal stabilizer readout, the collection of $g_j^Z$ outcomes identifies the realized bit-flip pattern on the returned probes.

The sensing statistics are unchanged because $X_B^{\boldsymbol{e}}$ commutes with every $g_j^X$:
\begin{equation}
p_{j,\mathrm{ret}}^{(\pm)}
=
p_j^{(\pm)},
\qquad
\langle g_j^X\rangle_{\mathrm{ret}}
=
\cos\theta_j .
\label{eq:return-sensing-statistics}
\end{equation}
The identified error pattern may be removed physically or tracked through a Pauli-frame update.

\section{Post-sensing logical-state preservation and reuse}\label{sec4}

The remote probes are physical subsystems of the entanglement-assisted encoded state rather than independently prepared sensing ancillas. The joint stabilizer readout must therefore extract the phase and return-error information without resolving the $k$ logical degrees of freedom carried by the same state.

The joint stabilizer measurements are assumed ideal and nondestructive, so that the conditional post-measurement state remains available after the classical outcomes are recorded. Since all $g_j^X$ and $g_j^Z$ commute, their measurement order is immaterial.

\begin{figure}[htb]
\centering
\includegraphics[width=0.9\textwidth]{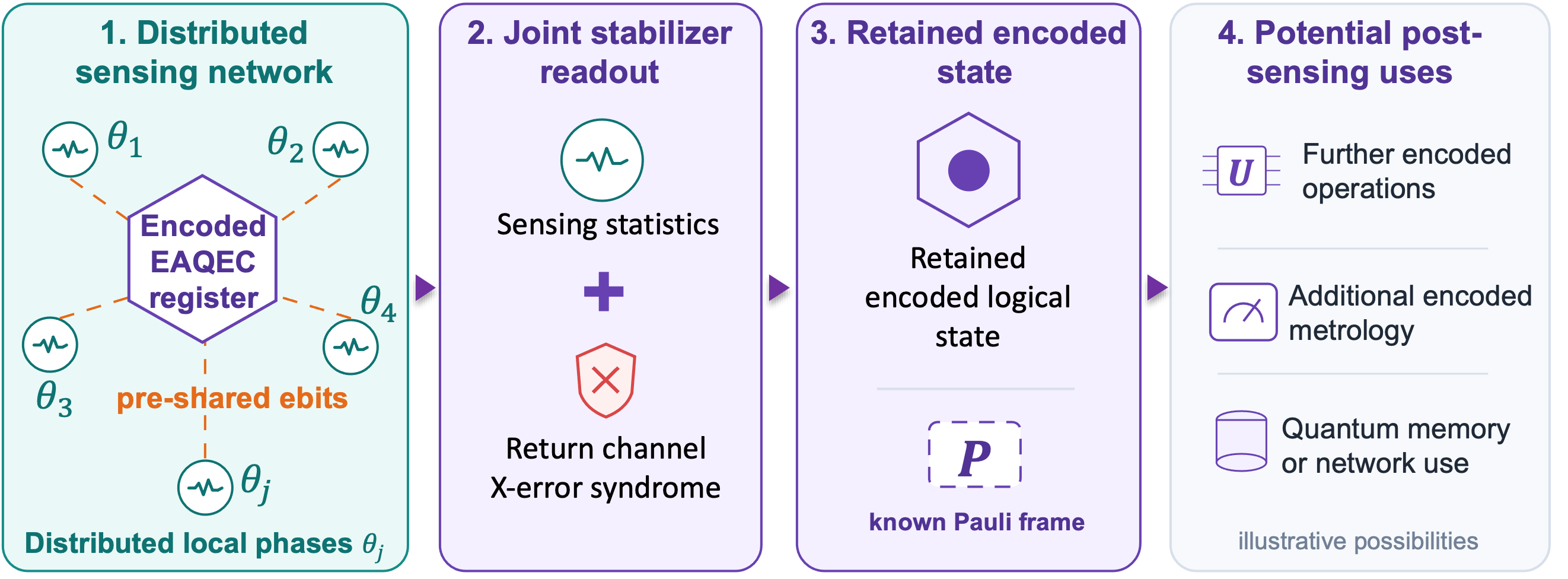}
\caption{Post-sensing reuse of the encoded state. After local phase acquisition, probe return, and joint stabilizer readout, the conditional encoded state differs from the initial state by a known Pauli transformation and remains available for subsequent encoded operations.}
\label{fig:scheme}
\end{figure}

Let $(\boldsymbol{s},\boldsymbol{e})$ denote the recorded sensing outcomes and return-error pattern. Logical-state preservation requires the normalized conditional state to have the form
\begin{equation}
\lvert\Psi_{\mathrm{post}}
(\boldsymbol{s},\boldsymbol{e};\psi)\rangle
=
R(\boldsymbol{s},\boldsymbol{e})
\lvert\Psi_{\mathrm{EA}}(\psi)\rangle,
\label{eq:logical-preservation-condition}
\end{equation}
where $R(\boldsymbol{s},\boldsymbol{e})$ is a known Pauli transformation determined by the measurement record and independent of the logical input $\lvert\psi\rangle$. The conditional state then occupies a known Pauli-shifted copy of the original code space and can be restored by physical correction or a Pauli-frame update. The resulting post-sensing reuse is illustrated in Fig.~\ref{fig:scheme}.

Write the sensing outcomes as $m_j^X=(-1)^{s_j}$, with $\boldsymbol{s}=(s_1,\ldots,s_c)\in\mathbb{F}_2^c$, and let $\boldsymbol{e}$ denote the return-error pattern identified from the $g_j^Z$ outcomes. The projector associated with the joint $g_j^X$ measurement record is
\begin{equation}
\Pi_{\boldsymbol{s}}^X
=
\prod_{j=1}^{c}
\frac{I+(-1)^{s_j}g_j^X}{2}.
\label{eq:sensing-projector}
\end{equation}

For any outcome record with nonzero probability,
\begin{equation}
\Pi_{\boldsymbol{s}}^X
X_B^{\boldsymbol{e}}
U_{\boldsymbol{\theta}}^{(B)}
\lvert\Psi_{\mathrm{EA}}(\psi)\rangle
\propto
X_B^{\boldsymbol{e}}
Z_B^{\boldsymbol{s}}
\lvert\Psi_{\mathrm{EA}}(\psi)\rangle,
\label{eq:conditional-projection}
\end{equation}
where $Z_B^{\boldsymbol{s}}
\equiv
\prod_{j=1}^{c}Z_{B_j}^{s_j}$. The phase-dependent coefficient determines the probability of the measurement record but is removed by normalization.

The state on the right-hand side is already an eigenstate of each $g_j^Z$ with eigenvalue $(-1)^{e_j}$; hence, the subsequent syndrome readout does not introduce an additional change to the conditional quantum state.

The conditional post-sensing state is therefore
\begin{equation}
\lvert\Psi_{\mathrm{post}}
(\boldsymbol{s},\boldsymbol{e};\psi)\rangle
=
X_B^{\boldsymbol{e}}
Z_B^{\boldsymbol{s}}
\lvert\Psi_{\mathrm{EA}}(\psi)\rangle
\label{eq:post-sensing-state}
\end{equation}
up to a global phase. Both $\boldsymbol{s}$ and $\boldsymbol{e}$ are known from the joint stabilizer record, so the Pauli transformation is fully determined and does not depend on the logical input.

Applying the known inverse Pauli $Z_B^{\boldsymbol{s}}X_B^{\boldsymbol{e}}$ restores $\lvert\Psi_{\mathrm{EA}}(\psi)\rangle$ up to a global phase. Because the correcting Pauli is independent of the logical input, the same recovery applies to arbitrary logical superpositions and extends by linearity to mixed logical states. The post-sensing state can therefore be retained in the encoded system and corrected physically or tracked through a Pauli frame.

\section{Estimation performance}
\label{sec5}

\subsection{Fisher-information characterization}\label{sec5-1}

For the binary $g_j^X$ measurement, $p_j^{(\pm)}=(1\pm\cos\theta_j)/2$. The corresponding classical Fisher information is
\begin{equation}
F_j^{(C)}(\theta_j)
=
\sum_{m=\pm}
\frac{1}{p_j^{(m)}}
\left(
\frac{\partial p_j^{(m)}}{\partial\theta_j}
\right)^2
=
1,
\qquad
0<\theta_j<\pi .
\label{eq:single-fi}
\end{equation}

The joint sensing-outcome distribution factorizes across the remote probes, with each outcome bit depending only on its corresponding local parameter. The classical Fisher-information matrix is therefore
\begin{equation}
\boldsymbol{F}^{(C)}(\boldsymbol{\theta})
=
I_c.
\label{eq:classical-fim}
\end{equation}
For $M$ independent repetitions, any locally unbiased estimator therefore satisfies
\begin{equation}
\operatorname{Cov}(\hat{\boldsymbol{\theta}})
\succeq
\frac{1}{M}I_c .
\label{eq:crb}
\end{equation}

For a pure encoded state, the generator of $\theta_j$ is $G_j=Z_{B_j}/2$, so that $F_{j\ell}^{(Q)}=4\,\operatorname{Cov}(G_j,G_\ell)$. Since the Bob-side reduced state is maximally mixed, $\langle Z_{B_j}\rangle=0$ and $\langle Z_{B_j}Z_{B_\ell}\rangle=\delta_{j\ell}$. Hence,
\begin{equation}
\boldsymbol{F}^{(C)}
=
\boldsymbol{F}^{(Q)}
=
I_c.
\label{eq:fi-equality}
\end{equation}

The stabilizer readout therefore extracts the quantum Fisher information available for the local-phase parameters in this construction. The unit Fisher information per parameter is the same as that of an optimally prepared single-qubit probe under the generator $Z/2$; the equality above indicates information-preserving readout rather than an intrinsic
metrological enhancement from the entanglement-assisted structure.

\subsection{Finite-shot estimation performance}\label{sec5-2}

\begin{figure}[ht!]
    \centering
    \includegraphics[width=0.72\linewidth]{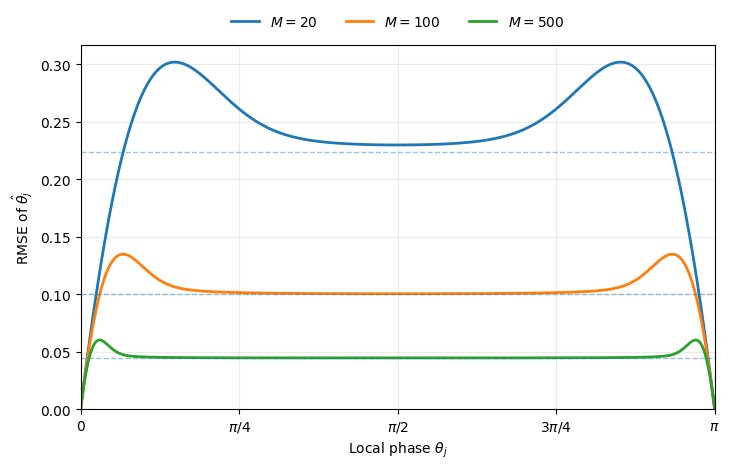}
    \caption{Exact finite-shot RMSE of the estimator $\hat{\theta}_j$ for $M=20$, $100$, and $500$, evaluated under the binomial measurement model. The horizontal dashed lines indicate the corresponding asymptotic levels $1/\sqrt{M}$.}
    \label{fig:result}
\end{figure}

For a fixed local phase $\theta_j$, suppose that the complete sensing procedure is repeated $M$ times. If $N_j^{(+)}$ denotes the number of $+1$ outcomes obtained from $g_j^X$, then $N_j^{(+)}\sim\operatorname{Binomial}(M,p_j^{(+)})$. With the observed frequency $\hat p_j^{(+)}=N_j^{(+)}/M$, the corresponding phase estimator is
\begin{equation}
\hat\theta_j
=
\arccos
\left(
2\frac{N_j^{(+)}}{M}-1
\right),
\qquad
\theta_j\in[0,\pi].
\label{eq:finite-shot-estimator}
\end{equation}

Because the inverse relation between $p_j^{(+)}$ and $\theta_j$ is nonlinear, the finite-shot error of this estimator does not generally coincide with its asymptotic Cramér--Rao behavior. For a true phase
$\theta_j$, its exact root-mean-square error is
\begin{equation}
\operatorname{RMSE}(\theta_j;M)
=
\left[
\sum_{n=0}^{M}
\binom{M}{n}
\left(p_j^{(+)}\right)^n
\left(1-p_j^{(+)}\right)^{M-n}
\left(
\arccos\left(2\frac{n}{M}-1\right)
-\theta_j
\right)^2
\right]^{1/2}.
\label{eq:exact-rmse}
\end{equation}

For phases in the interior of the identifiable interval, the estimator approaches the Fisher-information limit as $M$ increases,
\begin{equation}
\operatorname{Var}(\hat\theta_j)
\simeq
\frac{1}{M},
\qquad
\operatorname{RMSE}(\theta_j;M)
\simeq
\frac{1}{\sqrt{M}}.
\label{eq:asymptotic-rmse}
\end{equation}

Fig.~\ref{fig:result} shows the exact RMSE for $M=20$, $100$, and $500$. Over most of the interval, the RMSE approaches the corresponding $1/\sqrt{M}$ level as the number of repetitions increases. Deviations are concentrated near the interval boundaries, where the nonlinear inverse mapping amplifies the discreteness of the binomial measurement count. At the exact endpoints $\theta_j=0$ and $\pi$, the outcome is deterministic and the estimator returns the corresponding endpoint exactly, giving zero RMSE.

The same marginal estimation procedure applies to each local parameter. Under the independent bit-flip return channel considered here, the probabilities $p_j^{(\pm)}$ are independent of $q_j$ because the return errors commute with $g_j^X$. The return-error probability therefore does not enter the estimator or its RMSE, while the realized error pattern is recorded independently by the $g_j^Z$ outcomes.

%\subsection{Return-channel error statistics and sensing robustness}
%\label{subsec5-3}
%\subsection{Resource scaling and sensing capacity}
%\label{subsec5-3}

\section{Discussion}\label{sec6}

The role of error correction in the present construction differs from that in conventional quantum-error-corrected metrology. Error-corrected sensing schemes are typically designed to suppress detrimental noise while retaining sensitivity to the signal Hamiltonian, with the possibility of recovering coherence or favorable metrological scaling under suitable noise conditions \cite{Arrad2014,Kessler2014,Dur2014,Zhou2018}. The associated tension between protection and signal sensitivity has also motivated asymmetric code constructions that deliberately relax protection along the sensing direction \cite{Chen2025}. Here, the entanglement-assisted code is not introduced to suppress the local $Z$-phase signal or to enhance its sensitivity. Instead, the pre-existing bipartite stabilizer structure is used to support local phase acquisition while retaining the encoded logical subsystem. In this sense, the role of the code is primarily structural: the remote halves of the shared ebits, which ordinarily participate in the extended EAQEC stabilizer relations, are reassigned as spatially separated probes.

This setting also differs from the use of syndrome information purely for error diagnosis. Stabilizer syndromes can contain information about physical noise processes without requiring logical-state measurement \cite{Wagner2022}, and encoded signal-processing schemes have shown that syndrome measurements can themselves serve as sensing primitives \cite{OrtizMarrero2026}. In the present construction, the two operators associated with each remote ebit have complementary functions. The $X_{B_j}$-containing stabilizer is sensitive to the local $Z$-rotation, whereas its $Z_{B_j}$-containing partner remains insensitive to that signal and records the specified $X$-type return error. The distinction follows directly from the commutation relations of the remote stabilizer pair rather than from the introduction of a separate sensing register. At the same time, the logical degrees of freedom are not resolved by the joint readout: conditioned on the measurement record, the encoded state is displaced only by a known Pauli transformation. This combination distinguishes the construction from distributed-sensing schemes in which shared entanglement and error correction are introduced primarily to improve sensing performance across a sensor network \cite{Zhuang2020,ZhouBradyZhuang2022}.

The Fisher-information result should be interpreted in the same structural sense. The equality $\boldsymbol{F}^{(C)}=\boldsymbol{F}^{(Q)}=I_c$ shows that the stabilizer measurement extracts the local-phase information available in the encoded state. It does not imply that the entanglement-assisted structure increases the quantum Fisher information: under the generator $Z/2$, the available information per local parameter is the same as for an optimally prepared single-qubit probe. The finite-shot behavior is consistent with this interpretation, approaching the corresponding $1/\sqrt{M}$ scaling away from the boundaries of the identifiable interval. Its metrological role is therefore to preserve the available phase information while making the sensing readout compatible with return-error information and continued availability of the encoded logical state.

The present conclusions depend on several restrictions of the model. The pre-shared ebits, Alice-side encoding, and nondestructive stabilizer measurements are assumed ideal, and no additional noise is included on Alice's retained register during sensing and probe return. Noise on the returning probes is restricted to independent $X$ errors acting after phase acquisition; this restriction is essential to the separation between the sensing statistics carried by $g_j^X$ and the syndrome carried by $g_j^Z$. The conventional EAQEC distance $d$ continues to characterize errors on Alice's encoded register and does not provide a general protection guarantee for the remote probes. More general Bob-side errors may fail to commute with the sensing stabilizers and can modify both the phase-dependent statistics and the conditional post-sensing state. In addition, a single $g_j^X$ readout depends on the phase through $\cos\theta_j$, so direct phase identification requires an operating interval on which this response is one-to-one. It remains to determine whether the same sensing-and-reuse structure persists under imperfect stabilizer measurements, broader return-noise models, alternative signal generators, and other entanglement-assisted code families. A further question is whether code constructions can retain this structural compatibility while also providing genuine noise suppression or metrological gain.

\section{Conclusion}\label{sec7}

We introduced an entanglement-assisted stabilizer framework for distributed sensing of local phases, in which the remote halves of pre-shared ebits serve as spatially separated probes while the joint encoded state retains its logical degrees of freedom. For each remote probe, the $X_{B_j}$-containing stabilizer provides phase-dependent sensing statistics, whereas its $Z_{B_j}$-containing partner identifies the specified $X$-type return error. The graph-code construction makes this remote stabilizer structure explicit, with the $[[5,1,3;2]]$ code providing a concrete realization. Conditioned on the joint measurement record, the post-sensing state differs from the initial encoded state only by a known Pauli transformation and therefore remains available for subsequent encoded use.

For the local $Z$-phase model considered here, the stabilizer readout attains the quantum Fisher information available in the encoded state. The resulting framework thus establishes the compatibility of distributed local-phase sensing, restricted return-error identification, and post-sensing logical-state retention within a single entanglement-assisted stabilizer structure.

%\bmhead{Supplementary information}

%If your article has accompanying supplementary file/s please state so here.

%Authors reporting data from electrophoretic gels and blots should supply the full unprocessed scans for key as part of their Supplementary information. This may be requested by the editorial team/s if it is missing.

%Please refer to Journal-level guidance for any specific requirements.

\section*{Declarations}

\bmhead{Availability of data and materials}
Data sharing is not applicable to this article as no datasets were generated or analysed during the current study.

\bmhead{Competing interests}
The authors declare that they have no competing interests.

\bmhead{Funding}
This research was supported by the Korea Institute of Science and Technology Information (KISTI) (No.~(KISTI)K26L1M3C5-01).

\bmhead{Authors' contributions}
H.Z. and I.S. conceived the study. H.Z. developed the theoretical framework, performed the analysis, and drafted the manuscript. I.S. contributed to the development and verification of the theoretical framework and to discussions throughout the study. J.H. supervised the work and provided critical revision of the manuscript. All authors reviewed and approved the final manuscript.

\bmhead{Acknowledgements}
Not applicable.

\bmhead{Use of large language models}
ChatGPT (OpenAI) was used during manuscript preparation to assist with language editing, organization, and the drafting and checking of the mathematical exposition. All derivations, calculations, interpretations, and conclusions were reviewed and verified by the authors, who take full responsibility for the content of the manuscript.

\begin{appendices}

\section{Graph-code construction and remote-pair condition}
\label{Apen_A}

Consider the graph partition introduced in Sec.~\ref{sec2-2},
\begin{equation}
V=I\sqcup O,
\qquad
O=O_A\sqcup O_B,
\end{equation}
where $I$ contains the $k$ input vertices and $O$ contains the $n+c$
physical output vertices. Let $\Gamma$ denote the binary adjacency matrix
of the graph, and let $\Gamma_{S,T}$ denote the submatrix with rows indexed
by $S$ and columns indexed by $T$.

For an output vertex $v\in O$, define its neighborhood within the output
subgraph as
\begin{equation}
N_O(v)=N_G(v)\cap O.
\end{equation}
The associated graph-state vertex operator is
\begin{equation}
K_v
=
X_v
\prod_{u\in N_O(v)} Z_u .
\label{eq:app-vertex-operator}
\end{equation}
For a binary vector $x\in\mathbb{F}_2^{n+c}$, define
\begin{equation}
K(x)
=
\prod_{v\in O}K_v^{x_v}.
\label{eq:app-Kx}
\end{equation}
Up to an overall Pauli phase, this operator can be written as
\begin{equation}
K(x)
\doteq
X_O^x Z_O^{\Gamma_{O,O}x}.
\label{eq:app-binary-Kx}
\end{equation}

Let $|G_O\rangle$ denote the graph state of the output subgraph. The
encoded graph-basis states can be represented, up to a
logical-basis-dependent global phase, as
\begin{equation}
|\psi_y\rangle
\doteq
Z_O^{\Gamma_{O,I}y}|G_O\rangle,
\qquad
y\in\mathbb{F}_2^k .
\label{eq:app-graph-basis}
\end{equation}
Since $K(x)$ stabilizes $|G_O\rangle$, its action on an encoded
graph-basis state is
\begin{equation}
K(x)|\psi_y\rangle
=
(-1)^{y^T\Gamma_{I,O}x}
|\psi_y\rangle.
\label{eq:app-K-action}
\end{equation}
Therefore, $K(x)$ stabilizes every encoded basis state if and only if
\begin{equation}
\Gamma_{I,O}x=0.
\label{eq:app-kernel-condition}
\end{equation}
The graph-code stabilizer group is consequently
\begin{equation}
S_G
=
\left\{
K(x):
x\in\ker_{\mathbb{F}_2}(\Gamma_{I,O})
\right\}.
\label{eq:app-graph-stabilizer}
\end{equation}
For a valid $k$-input graph code with $\operatorname{rank}(\Gamma_{I,O})=k$, the kernel has dimension $n+c-k$, and the corresponding stabilized subspace has dimension $2^k$
\cite{SchlingemannWerner2001,Schlingemann2002}.

We now apply this characterization to the remote-leaf-matched structure. For each remote vertex $B_j\in O_B$,
\begin{equation}
N_G(B_j)=\{a_j\},
\qquad
a_j\in O_A.
\label{eq:app-remote-leaf}
\end{equation}
Because $B_j$ has no input neighbor, the binary vector selecting only $B_j$ lies in $\ker(\Gamma_{I,O})$. Its vertex operator therefore belongs to $S_G$ and is
\begin{equation}
K_{B_j}
=
Z_{a_j}X_{B_j}.
\label{eq:app-remote-K}
\end{equation}
This gives the remote stabilizer
\begin{equation}
g_j^X
=
K_{B_j}
=
Z_{a_j}X_{B_j}.
\label{eq:app-remote-gx}
\end{equation}

It remains to determine when the stabilizer group contains a compatible partner whose Bob-side restriction is $Z_{B_j}$. For a remote-leaf-matched graph, $B_\ell$ is adjacent only to $a_\ell$, so that
\begin{equation}
(\Gamma_{O,O}x)_{B_\ell}
=
x_{a_\ell}.
\label{eq:app-bob-z-support}
\end{equation}
A stabilizer with Bob-side restriction $Z_{B_j}$ exists if there is a vector $x^{(j)}$ satisfying
\begin{equation}
\Gamma_{I,O}x^{(j)}=0,
\qquad
x^{(j)}\big|_{O_B}=0,
\qquad
x^{(j)}_{a_\ell}=\delta_{j\ell}.
\label{eq:app-gz-condition}
\end{equation}
The second condition removes $X$-type support from Bob's register, while the third restricts its $Z$-type support to $B_j$. Consequently,
\begin{equation}
K\!\left(x^{(j)}\right)\big|_{O_B}
=
Z_{B_j},
\label{eq:app-gz-bob}
\end{equation}
and we may identify
\begin{equation}
g_j^Z
=
K\!\left(x^{(j)}\right)
=
Q_j\otimes Z_{B_j}.
\label{eq:app-remote-gz}
\end{equation}
Since $x^{(j)}_{a_j}=1$, the Alice-side restriction $Q_j$ contains $X$ or $Y$ on $a_j$ and therefore anticommutes with the $Z_{a_j}$ factor of $g_j^X$. This Alice-side anticommutation is compensated by the anticommutation of $X_{B_j}$ and $Z_{B_j}$, yielding the commuting joint pair required by the entanglement-assisted stabilizer structure.

\setcounter{equation}{0}

\section{Explicit $[[5,1,3;2]]$ entanglement-assisted graph-code construction}
\label{Apen_B}

We verify here the $[[5,1,3;2]]$ realization shown in Fig.~\ref{fig:example}. Its vertex sets are
\begin{equation}
I=\{i\},\qquad
O_A=\{A_1,A_2,A_3,A_4,A_5\},\qquad
O_B=\{B_1,B_2\}.
\end{equation}

The Alice-side output vertices form the cycle
\begin{equation}
A_1-A_2-A_3-A_4-A_5-A_1,
\end{equation}
with additional remote edges $A_4-B_1$ and $A_5-B_2$. The input vertex $i$ is adjacent to all five Alice-side output vertices and to neither remote vertex. With the output ordering
\begin{equation}
(A_1,A_2,A_3,A_4,A_5,B_1,B_2),
\end{equation}
the input-output adjacency block is
\begin{equation}
\Gamma_{I,O}
=
\begin{pmatrix}
1&1&1&1&1&0&0
\end{pmatrix}.
\label{eq:B1}
\end{equation}

According to the kernel characterization in Appendix~A, a binary vector
$x\in\mathbb{F}_2^7$ generates a graph-code stabilizer whenever
\begin{equation}
x_{A_1}+x_{A_2}+x_{A_3}+x_{A_4}+x_{A_5}
=
0
\pmod 2.
\label{eq:B2}
\end{equation}
Since $\operatorname{rank}(\Gamma_{I,O})=1$, the kernel has dimension
six, consistent with a one-logical-qubit code on the seven-qubit joint
output register.

The output-subgraph vertex operators are
\begin{align}
K_{A_1} &= X_{A_1}Z_{A_2}Z_{A_5},
&
K_{A_2} &= Z_{A_1}X_{A_2}Z_{A_3},
\nonumber\\
K_{A_3} &= Z_{A_2}X_{A_3}Z_{A_4},
&
K_{A_4} &= Z_{A_3}X_{A_4}Z_{A_5}Z_{B_1},
\nonumber\\
K_{A_5} &= Z_{A_1}Z_{A_4}X_{A_5}Z_{B_2},
&
K_{B_1} &= Z_{A_4}X_{B_1},
\nonumber\\
&&
K_{B_2} &= Z_{A_5}X_{B_2}.
\label{eq:B3}
\end{align}

A convenient generating set satisfying the kernel condition is
\begin{align}
h_1
&=
K_{A_1}K_{A_2}
=
Y_{A_1}Y_{A_2}Z_{A_3}Z_{A_5},
\nonumber\\
h_2
&=
K_{A_2}K_{A_3}
=
Z_{A_1}Y_{A_2}Y_{A_3}Z_{A_4},
\nonumber\\
g_1^X
&=
K_{B_1}
=
Z_{A_4}X_{B_1},
\nonumber\\
g_1^Z
&=
K_{A_1}K_{A_4}
=
X_{A_1}Z_{A_2}Z_{A_3}X_{A_4}Z_{B_1},
\nonumber\\
g_2^X
&=
K_{B_2}
=
Z_{A_5}X_{B_2},
\nonumber\\
g_2^Z
&=
K_{A_2}K_{A_5}
=
X_{A_2}Z_{A_3}Z_{A_4}X_{A_5}Z_{B_2}.
\label{eq:B4}
\end{align}

These six operators mutually commute, and a direct binary-rank check shows
that they are independent. They therefore stabilize a subspace of
dimension
\begin{equation}
2^{7-6}=2,
\label{eq:B5}
\end{equation}
corresponding to one encoded logical qubit. In particular, the two remote
pairs have the form required in Sec.~\ref{sec2-1}: $g_j^X$ contains
$X_{B_j}$, while $g_j^Z$ contains $Z_{B_j}$, with the corresponding
Alice-side restrictions anticommuting.

One possible pair of logical Pauli representatives is
\begin{equation}
\bar{X}
=
X_{A_1}X_{A_2}Z_{A_4},
\qquad
\bar{Z}
=
Z_{A_1}X_{A_2}Z_{A_3}.
\label{eq:B6}
\end{equation}
Both commute with all six stabilizer generators, while
\begin{equation}
\bar{X}\bar{Z}
=
-\bar{Z}\bar{X},
\label{eq:B7}
\end{equation}
and therefore represent the logical $X$ and $Z$ operators of the encoded
qubit.

A direct enumeration of Pauli operators supported on Alice's five-qubit
register shows that no operator of weight one or two commutes with all six
stabilizer generators while acting nontrivially on the code space. Since
the logical representatives above have Alice-side weight three, the
minimum conventional logical weight is three. The resulting
entanglement-assisted code therefore has parameters $[[5,1,3;2]]$.

\setcounter{equation}{0}

\section{Conditional-state derivation}
\label{Apen_C}

Let
\begin{equation}
\lvert\Psi_0\rangle
=
\lvert\Psi_{\mathrm{EA}}(\psi)\rangle .
\end{equation}

The multi-probe sensing unitary can be expanded as
\begin{equation}
U_{\boldsymbol{\theta}}^{(B)}
=
\sum_{\boldsymbol{r}\in\mathbb{F}_2^c}
a_{\boldsymbol{r}}(\boldsymbol{\theta})
Z_B^{\boldsymbol{r}},
\label{eq:C1}
\end{equation}
where
\begin{equation}
a_{\boldsymbol{r}}(\boldsymbol{\theta})
=
\prod_{j=1}^{c}
\left(
\cos\frac{\theta_j}{2}
\right)^{1-r_j}
\left(
-i\sin\frac{\theta_j}{2}
\right)^{r_j},
\qquad
Z_B^{\boldsymbol{r}}
=
\prod_{j=1}^{c}Z_{B_j}^{r_j}.
\label{eq:C2}
\end{equation}

Since $g_j^X\lvert\Psi_0\rangle=\lvert\Psi_0\rangle$, while $Z_{B_j}$ anticommutes only with the $X_{B_j}$ factor of the corresponding sensing stabilizer,
\begin{equation}
g_j^X
Z_B^{\boldsymbol{r}}
\lvert\Psi_0\rangle
=
(-1)^{r_j}
Z_B^{\boldsymbol{r}}
\lvert\Psi_0\rangle .
\label{eq:C3}
\end{equation}
Thus, $Z_B^{\boldsymbol{r}}\lvert\Psi_0\rangle$ belongs to the joint sensing sector labelled by $\boldsymbol{r}$. Projection onto the observed sector $\boldsymbol{s}$ gives
\begin{equation}
\Pi_{\boldsymbol{s}}^X
U_{\boldsymbol{\theta}}^{(B)}
\lvert\Psi_0\rangle
=
a_{\boldsymbol{s}}(\boldsymbol{\theta})
Z_B^{\boldsymbol{s}}
\lvert\Psi_0\rangle .
\label{eq:C4}
\end{equation}

The corresponding joint sensing probability is
\begin{equation}
p(\boldsymbol{s}\mid\boldsymbol{\theta})
=
\left|a_{\boldsymbol{s}}(\boldsymbol{\theta})\right|^2
=
\prod_{j=1}^{c}
\left(
\cos^2\frac{\theta_j}{2}
\right)^{1-s_j}
\left(
\sin^2\frac{\theta_j}{2}
\right)^{s_j}.
\label{eq:C5}
\end{equation}

For a realized return-error pattern $\boldsymbol{e}$, $X_B^{\boldsymbol{e}}$ commutes with every sensing stabilizer, so
\begin{equation}
\Pi_{\boldsymbol{s}}^X
X_B^{\boldsymbol{e}}
U_{\boldsymbol{\theta}}^{(B)}
\lvert\Psi_0\rangle
=
a_{\boldsymbol{s}}(\boldsymbol{\theta})
X_B^{\boldsymbol{e}}
Z_B^{\boldsymbol{s}}
\lvert\Psi_0\rangle .
\label{eq:C6}
\end{equation}

Moreover,
\begin{equation}
g_j^Z
X_B^{\boldsymbol{e}}
Z_B^{\boldsymbol{s}}
\lvert\Psi_0\rangle
=
(-1)^{e_j}
X_B^{\boldsymbol{e}}
Z_B^{\boldsymbol{s}}
\lvert\Psi_0\rangle ,
\label{eq:C7}
\end{equation}
so the $g_j^Z$ outcomes identify $\boldsymbol{e}$. Normalization then gives the conditional post-sensing state in Eq.~(\ref{eq:post-sensing-state}).

\end{appendices}

%\bibliography{references}

\begin{thebibliography}{10}
\providecommand{\doi}[1]{\url{https://doi.org/#1}}
\bibcommenthead

\bibitem[\protect\citeauthoryear{Giovannetti et~al.}{2011}]{Giovannetti2011}
Giovannetti V, Lloyd S, Maccone L.
\newblock Advances in quantum metrology.
\newblock Nature Photonics. 2011 March;5(4):222--229.
\newblock \doi{10.1038/nphoton.2011.35}.

\bibitem[\protect\citeauthoryear{Proctor et~al.}{2018}]{Proctor2018}
Proctor TJ, Knott PA, Dunningham JA.
\newblock Multiparameter Estimation in Networked Quantum Sensors.
\newblock Physical Review Letters. 2018 February;120(8):080501.
\newblock \doi{10.1103/PhysRevLett.120.080501}.

\bibitem[\protect\citeauthoryear{Eldredge et~al.}{2018}]{Eldredge2018}
Eldredge Z, Foss-Feig M, Gross JA, Rolston SL, Gorshkov AV.
\newblock Optimal and secure measurement protocols for quantum sensor networks.
\newblock Physical Review A. 2018 April;97(4):042337.
\newblock \doi{10.1103/PhysRevA.97.042337}.

\bibitem[\protect\citeauthoryear{Kim et~al.}{2024}]{Kim2024}
Kim DH, Hong S, Kim YS, Kim Y, Lee SW, Pooser RC, et~al.
\newblock Distributed quantum sensing of multiple phases with fewer photons.
\newblock Nature Communications. 2024;15:266.
\newblock \doi{10.1038/s41467-023-44204-z}.

\bibitem[\protect\citeauthoryear{Bate et~al.}{2025}]{Bate2025}
Bate J, Hamann A, Canteri M, Winkler A, Koong ZX, Krutyanskiy V, et~al.
\newblock Experimental Distributed Quantum Sensing in a Noisy Environment.
\newblock Physical Review Letters. 2025;135(22):220801.
\newblock \doi{10.1103/3hgx-wcdn}.

\bibitem[\protect\citeauthoryear{Zhang et~al.}{2026}]{Zhang2026}
Zhang J, Wang L, Hai YJ, Zhang J, Chu J, Jiang J, et~al.
\newblock Distributed multi-parameter quantum metrology with a superconducting
  quantum network.
\newblock Nature Communications. 2026;17:1825.
\newblock \doi{10.1038/s41467-026-68535-9}.

\bibitem[\protect\citeauthoryear{Arrad et~al.}{2014}]{Arrad2014}
Arrad G, Vinkler Y, Aharonov D, Retzker A.
\newblock Increasing Sensing Resolution with Error Correction.
\newblock Physical Review Letters. 2014;112(15):150801.
\newblock \doi{10.1103/PhysRevLett.112.150801}.

\bibitem[\protect\citeauthoryear{Kessler et~al.}{2014}]{Kessler2014}
Kessler EM, Lovchinsky I, Sushkov AO, Lukin MD.
\newblock Quantum Error Correction for Metrology.
\newblock Physical Review Letters. 2014;112(15):150802.
\newblock \doi{10.1103/PhysRevLett.112.150802}.

\bibitem[\protect\citeauthoryear{D{\"u}r et~al.}{2014}]{Dur2014}
D{\"u}r W, Skotiniotis M, Fr{\"o}wis F, Kraus B.
\newblock Improved Quantum Metrology Using Quantum Error Correction.
\newblock Physical Review Letters. 2014;112(8):080801.
\newblock \doi{10.1103/PhysRevLett.112.080801}.

\bibitem[\protect\citeauthoryear{Zhou et~al.}{2018}]{Zhou2018}
Zhou S, Zhang M, Preskill J, Jiang L.
\newblock Achieving the Heisenberg Limit in Quantum Metrology Using Quantum
  Error Correction.
\newblock Nature Communications. 2018;9:78.
\newblock \doi{10.1038/s41467-017-02510-3}.

\bibitem[\protect\citeauthoryear{Mann et~al.}{2025}]{Mann2025}
Mann Z, Cao N, Laflamme R, Zhou S.
\newblock Quantum Error-Corrected Non-Markovian Metrology.
\newblock PRX Quantum. 2025;6(3):030321.
\newblock \doi{10.1103/wfyl-wtz3}.

\bibitem[\protect\citeauthoryear{Layden et~al.}{2019}]{Layden2019}
Layden D, Zhou S, Cappellaro P, Jiang L.
\newblock Ancilla-Free Quantum Error Correction Codes for Quantum Metrology.
\newblock Physical Review Letters. 2019;122(4):040502.
\newblock \doi{10.1103/PhysRevLett.122.040502}.

\bibitem[\protect\citeauthoryear{Zhou et~al.}{2024}]{Zhou2024}
Zhou S, Giannisis Manes A, Jiang L.
\newblock Achieving metrological limits using ancilla-free quantum
  error-correcting codes.
\newblock Physical Review A. 2024;109(4):042406.
\newblock \doi{10.1103/PhysRevA.109.042406}.

\bibitem[\protect\citeauthoryear{Chen et~al.}{2025}]{Chen2025}
Chen J, Luo R, Du Z, Yan Y, Zhou Y, Ma X.
\newblock Bypassing the protection-sensitivity incompatibility in
  quantum-error-corrected metrology via asymmetric codes.
\newblock arXiv:2512.20426 [quant-ph]. 2025.
\newblock \doi{10.48550/arXiv.2512.20426}.

\bibitem[\protect\citeauthoryear{Wagner et~al.}{2022}]{Wagner2022}
Wagner T, Kampermann H, Bru{\ss} D, Kliesch M.
\newblock Pauli Channels Can Be Estimated from Syndrome Measurements in Quantum
  Error Correction.
\newblock Quantum. 2022;6:809.
\newblock \doi{10.22331/q-2022-09-19-809}.

\bibitem[\protect\citeauthoryear{Ortiz Marrero et~al.}{2026}]{OrtizMarrero2026}
Ortiz Marrero C, Tang RJ, Wiebe N.
\newblock Encoded Quantum Signal Processing for Heisenberg-Limited Metrology.
\newblock arXiv:2603.22798 [quant-ph]. 2026.
\newblock \doi{10.48550/arXiv.2603.22798}.

\bibitem[\protect\citeauthoryear{Brun et~al.}{2006}]{Brun2006}
Brun TA, Devetak I, Hsieh MH.
\newblock Correcting Quantum Errors with Entanglement.
\newblock Science. 2006;314(5798):436--439.
\newblock \doi{10.1126/science.1131563}.

\bibitem[\protect\citeauthoryear{Hsieh et~al.}{2007}]{Hsieh2007}
Hsieh MH, Devetak I, Brun TA.
\newblock General Entanglement-Assisted Quantum Error-Correcting Codes.
\newblock Physical Review A. 2007;76(6):062313.
\newblock \doi{10.1103/PhysRevA.76.062313}.

\bibitem[\protect\citeauthoryear{Lai and Brun}{2012}]{LaiBrun2012}
Lai CY, Brun TA.
\newblock Entanglement-Assisted Quantum Error-Correcting Codes with Imperfect
  Ebits.
\newblock Physical Review A. 2012;86(3):032319.
\newblock \doi{10.1103/PhysRevA.86.032319}.

\bibitem[\protect\citeauthoryear{Zhuang et~al.}{2020}]{Zhuang2020}
Zhuang Q, Preskill J, Jiang L.
\newblock Distributed Quantum Sensing Enhanced by Continuous-Variable Error
  Correction.
\newblock New Journal of Physics. 2020;22(2):022001.
\newblock \doi{10.1088/1367-2630/ab7257}.

\bibitem[\protect\citeauthoryear{Zhou et~al.}{2022}]{ZhouBradyZhuang2022}
Zhou B, Brady AJ, Zhuang Q.
\newblock Enhancing Distributed Sensing with Imperfect Error Correction.
\newblock Physical Review A. 2022;106(1):012404.
\newblock \doi{10.1103/PhysRevA.106.012404}.

\bibitem[\protect\citeauthoryear{Schlingemann and
  Werner}{2001}]{SchlingemannWerner2001}
Schlingemann D, Werner RF.
\newblock Quantum Error-Correcting Codes Associated with Graphs.
\newblock Physical Review A. 2001;65(1):012308.
\newblock \doi{10.1103/PhysRevA.65.012308}.

\bibitem[\protect\citeauthoryear{Schlingemann}{2002}]{Schlingemann2002}
Schlingemann D.
\newblock Stabilizer Codes Can Be Realized as Graph Codes.
\newblock Quantum Information and Computation. 2002;2(4):307--323.
\newblock \doi{10.26421/QIC2.4-4}.
\newblock
  {\href{https://arxiv.org/abs/quant-ph/0111080}{{arXiv:quant-ph/0111080}}}.

\end{thebibliography}

\end{document}